\documentclass[11pt]{JHEP3}
\usepackage{latexsym}

\def\d{\delta}

\def\half{\frac{1}{2}}

\def\s{\sigma}

\def\we{\wedge}

\def\ra{\rightarrow}

\def\az{A_z}
\def\azb{A_{\bar{z}}}
\def\abz{\bar{A}_z}
\def\abzb{\bar{A}_{\bar{z}}}

\renewcommand{\@}[1]{\sqrt{#1}}

\def\be{\begin{eqnarray}}
\renewcommand{\le}[1]{\label{#1}\end{eqnarray}}
\def\ee{\end{eqnarray}}

\def\ffract#1#2{\raise .35 em\hbox{$\scriptstyle#1$}\kern-.25em/
\kern-.2em\lower .22 em \hbox{$\scriptstyle#2$}}

\def\del{\partial}

\def\half{{1\over2}}

\expandafter\ifx\csname amssym.def\endcsname\relax \else\endinput\fi
\expandafter\edef\csname amssym.def\endcsname{%
       \catcode`\noexpand\@=\the\catcode`\@\space}
\catcode`\@=11
\def\undefine#1{\let#1\undefined}
\def\newsymbol#1#2#3#4#5{\let\next@\relax
 \ifnum#2=\@ne\let\next@\msafam@\else
 \ifnum#2=\tw@\let\next@\msbfam@\fi\fi
 \mathchardef#1="#3\next@#4#5}
\def\mathhexbox@#1#2#3{\relax
 \ifmmode\mathpalette{}{\m@th\mathchar"#1#2#3}%
 \else\leavevmode\hbox{$\m@th\mathchar"#1#2#3$}\fi}
\def\hexnumber@#1{\ifcase#1 0\or 1\or 2\or 3\or 4\or 5\or 6\or 7\or 8\or
 9\or A\or B\or C\or D\or E\or F\fi}

\font\tenmsa=msam10
\font\sevenmsa=msam7
\font\fivemsa=msam5
\newfam\msafam
\textfont\msafam=\tenmsa
\scriptfont\msafam=\sevenmsa
\scriptscriptfont\msafam=\fivemsa
\edef\msafam@{\hexnumber@\msafam}
\mathchardef\dabar@"0\msafam@39
\def\dashrightarrow{\mathrel{\dabar@\dabar@\mathchar"0\msafam@4B}}
\def\dashleftarrow{\mathrel{\mathchar"0\msafam@4C\dabar@\dabar@}}

\def\ulcorner{\delimiter"4\msafam@70\msafam@70 }
\def\urcorner{\delimiter"5\msafam@71\msafam@71 }
\def\llcorner{\delimiter"4\msafam@78\msafam@78 }
\def\lrcorner{\delimiter"5\msafam@79\msafam@79 }
\def\yen{{\mathhexbox@\msafam@55}}
\def\checkmark{{\mathhexbox@\msafam@58}}
\def\circledR{{\mathhexbox@\msafam@72}}
\def\maltese{{\mathhexbox@\msafam@7A}}
\def\circledS{{\mathhexbox@\msafam@73}}

\font\tenmsb=msbm10
\font\sevenmsb=msbm7
\font\fivemsb=msbm5
\newfam\msbfam
\textfont\msbfam=\tenmsb
\scriptfont\msbfam=\sevenmsb
\scriptscriptfont\msbfam=\fivemsb
\edef\msbfam@{\hexnumber@\msbfam}
\def\Bbb#1{{\fam\msbfam\relax#1}}
\def\widehat#1{\setbox\z@\hbox{$\m@th#1$}%
 \ifdim\wd\z@>\tw@ em\mathaccent"0\msbfam@5B{#1}%
 \else\mathaccent"0362{#1}\fi}
\def\widetilde#1{\setbox\z@\hbox{$\m@th#1$}%
 \ifdim\wd\z@>\tw@ em\mathaccent"0\msbfam@5D{#1}%
 \else\mathaccent"0365{#1}\fi}
\font\teneufm=eufm10
\font\seveneufm=eufm7
\font\fiveeufm=eufm5
\newfam\eufmfam
\textfont\eufmfam=\teneufm
\scriptfont\eufmfam=\seveneufm
\scriptscriptfont\eufmfam=\fiveeufm
\def\frak#1{{\fam\eufmfam\relax#1}}

\csname amssym.def\endcsname

\renewcommand{\nc}{\newcommand}
\newcommand{\rnc}{\renewcommand}
\nc{\CC}{\Bbb{C}}
\nc{\HH}{\Bbb{H}}
\nc{\PP}{\Bbb{P}}
\nc{\RR}{\Bbb{R}}
\nc{\ZZ}{\Bbb{Z}}
\nc{\II}{\Bbb{I}}
\nc{\EE}{\Bbb{E}}
\rnc{\lg}{\frak{g}}
\nc{\lgc}{\lg_{\CC}}
\nc{\gc}{G_{\CC}}
\nc{\G}{{\cal G}}
\nc{\GC}{\G_{\CC}}
\nc{\zb}{{\bar{z}}}

\newcommand{\ex}[1]{\mbox{e}^{\,\textstyle#1}}
\newcommand{\tr}{\mathop{\mbox{tr}}\nolimits}

\title{On the Boundary Dynamics of Chern-Simons Gravity}
\author{Giovanni Arcioni\\Spinoza Institute and Institute for Theoretical
Physics, Utrecht University,\\ Leuvenlaan 4, 3584 CE Utrecht, The Netherlands\\
\email{G.Arcioni@phys.uu.nl}}
\author{Matthias Blau\\ICTP, Strada Costiera 11, I--34014 Trieste, Italy\\
\email{mblau@ictp.trieste.it}}
\author{Martin O'Loughlin\\S.I.S.S.A. Scuola Internazionale Superiore di Studi
Avanzati,\\ Via Beirut 4, I--34014 Trieste, Italy\\
\email{loughlin@sissa.it}}

\abstract{We study Chern-Simons theory with a complex $\gc$ or a real $G \times
G$ gauge group on a manifold with boundary - this includes Lorentzian
and Euclidean (anti-) de Sitter (E/A)dS gravity for $G=SU(2)$ or
$G=SL(2,\RR)$.  We show that there is a canonical choice of boundary
conditions that leads to an unambiguous, fully covariant and gauge
invariant, off-shell derivation of the boundary action - a $\gc/G$
or $G$ WZW model, coupled in a gauge invariant way to the boundary 
value of the gauge field. In particular, for (E/A)dS gravity, the boundary
action is a WZW model with target space $(E/A)dS_3$, reminiscent of a
worldsheet for worldsheet mechanism.  We discuss in some detail the
properties of the boundary theories that  arise and we confront our
results with various related constructions in the literature.}

\preprint{SISSA/65/2002/EP\\SPIN-2002/16\\ITF-2002/25\\hep-th/0210089}


\begin{document}


\newpage
\setcounter{footnote}{0}


\section{Introduction and Discussion}

In this paper we revisit the issue of `boundary dynamics' in
three-dimensional gravity. The main goal is to uncover the universal
nature of a natural set of boundary conditions and to give a completely
general, covariant and off-shell derivation of the corresponding boundary
action. In particular, we find that the target space of the boundary theory
sigma model is equal to the model geometry of the bulk gravity theory, e.g.\
the boundary action of asymptotically de Sitter gravity is a Wess-Zumino-Witten
model with target space three-dimensional de Sitter space $dS_3$, etc. Furthermore,
we are able to give a somewhat unified discussion of other results in this field.

It is well known \cite{bh} that asymptotically
Anti-de Sitter (AdS) gravity has an infinite-dimensional algebra of
asymptotic symmetries which is a Virasoro algebra with a non-trivial
classical central charge 
(and likewise for de Sitter (dS) space \cite{park1,dscft1}). 
This fits in neatly with the $AdS_3/CFT_2$ correspondence
based on string theory on $AdS_3$ and its holographically dual CFT -
see e.g. \cite{GKS,BORT,JMHO}. 

On the other hand, three-dimensional gravity can be formulated
\cite{2+1witten,at,carlip} as a Chern-Simons gauge theory
\cite{wittencs},\footnote{See \cite{banados1,park2} for a derivation of the
asymptotic Virasoro algebra in this context.} and this provides an
({\em a priori} different) relation between (A)dS gravity and CFT,
manifested e.g.\ in the fact that Chern-Simons theory (quite generally)
induces a chiral Wess-Zumino-Witten (WZW) model on the boundary
\cite{EMSS}. Beginning with the work of \cite{chvd}, attempts have been
made to attribute a dynamical origin to this conformal symmetry purely
within three-dimensional gravity, by relating this Virasoro algebra, along
the lines of \cite{EMSS}, to a boundary conformal field theory obtained by
a suitable restriction of the (topological) bulk gravitational dynamics
to the boundary.  This `boundary dynamics' is typically a Liouville
theory obtained as a constrained WZW model from the on-shell value of
the gravitational action, the constraints arising from the asymptotic
boundary conditions.

This boundary Liouville action undoubtedly captures certain aspects of the
gravitational dynamics - see e.g.\ \cite{ksss,bers,kk0,kv} for discussions in
the gravity context and \cite{d1d5} for the emergence of Liouville theory
in the so-called long string sector of string theory on $AdS_3$. 
But the overall significance of this boundary action for
three-dimensional quantum gravity has remained somewhat unclear. In
particular, there appears to be some controversy between the general
relativity and string theory camps regarding the relation of the
boundary action to a holographic description of three-dimensional
gravity. For a clear discussion of these issues from a string
theory perspective see \cite{martinec} - for a different point of view 
see e.g.\ the discussion in \cite{kk2}.

We believe that this unsatisfactory state of affairs is in part due to the
fact that the standard derivations of the boundary action, as described in
\cite{chvd} for AdS and in \cite{ck} for dS gravity, are rather
non-covariant (requiring detailed assumptions about the asymptotics
of the fields, the boundary, special choices of coordinates and/or
gauge conditions), and make use of the bulk (and boundary) equations
of motion. The latter, in particular, makes it difficult to assess the
significance of the boundary action for three-dimensional quantum gravity.

In this note we provide
a complementary approach to the the standard derivations of the
boundary action. We will study Chern-Simons theory with a complex $\gc$
\cite{cscwitten} or a real $G \times G$ gauge group on an arbitrary
three-manifold with boundary. This includes Lorentzian and Euclidean
(anti-) de Sitter (E/A)dS gravity, with $G=SU(2)$ or $G=SU(1,1)\sim
SL(2,\RR)$ the `Lorentz' group of local frame rotations.  We will show
that there is a canonical choice of boundary terms, fixed uniquely
by the requirement that the complete bulk plus boundary action be
$G$-invariant plus the requirement that half of the gauge fields are
fixed on the boundary.  In particular, in the gravitational context
the former is just the requirement of invariance under local frame
rotations, a natural requirement as this gauge symmetry is an artefact
of the dreibein formulation of gravity, while the latter is compatible
with the structure of asymptotically (E/A)dS gravity.

We will show that this choice of boundary terms leads to an unambiguous,
fully covariant and gauge invariant, {\em off-shell} derivation of
the boundary action, which turns out to be a $\gc/G$ or $G$ WZW model,
coupled in a gauge invariant way to the boundary value of the gauge
fields which act as sources for the currents of the WZW model. This
coupling is such that, if these boundary gauge fields were to be treated
as dynamical fields (in our approach they are not), then these actions
would describe topological field theories, either the standard $G/G$
models \cite{gmodg} or novel ``$(\gc/G)/G$'' models, presumably related
to the fusion rules of the $\gc/G$ WZW model.

In particular, for (E/A)dS gravity, in each case the boundary action is
a WZW model with target space the space(time) $(E/A)dS_3$ itself, i.e.\
the string theory sigma model with the NS target space B-field required
by conformal invariance. E.g.\ for $AdS_3$ this is a WZW model with
target space $SU(1,1)\sim SL(2,\RR)$, and this reproduces the result of
\cite{chvd}, where the $SL(2,\RR)$ WZW model appears at an intermediate
step in the derivation of Liouville theory. But we believe that in their present
generality the results are new, and that our general derivation provides
a novel perspective and clarifies the role of the boundary action.

For example, we do not see any obvious decoupling of the bulk and
boundary theories for quantities other than the bulk partition function,
and our derivation shows very clearly that the complete bulk gravitational
dynamics (including bulk correlation functions) is not captured by the
boundary dynamics alone but by the combined bulk plus boundary action. We
are thus not making any claims about a potentially dual two-dimensional
field theory and about the holographic degrees of freedom.

Our result is also reminiscent of a worldsheet for worldsheet mechanism
\cite{greenws,n2}. To further develop and strengthen this possible
analogy between the target space of our boundary action and the base space
of our quantum gravity theory one would need to study the Wilson lines of
Chern-Simons gravity and the corresponding boundary operator insertions.

We also want to emphasise that the present derivation
circumvents what is perhaps the least attractive feature of the standard
derivations - the necessity to combine two chiral WZW models into a
non-chiral WZW model by using the boundary equations of motion.\footnote{In the case 
of dS gravity \cite{ck} this procedure is moreover ambiguous, leading 
to either an $SL(2,\CC)/SU(2)$ or an $SL(2,\CC)/SU(1,1)$ WZW model - clearly very 
different theories.}
In our derivation, this recombination is automatic and off-shell due to the
Polyakov-Wiegman identities (the relevant Polyakov-Wiegman term being
provided by the boundary term which ensures $G$-invariance of the action).

We close this Introduction with some comments on other related work.

Of all the articles on the boundary dynamics of three-dimensional gravity,
\cite{kk1} is perhaps the one closest in spirit to the present work. We
will discuss the relation between \cite{kk1} and our work in Section 2.4. 
In \cite{banadosritz} (see \cite{banadosothers} for related work) 
the $SL(2,\CC)/SU(2)$ boundary WZW model is also interpreted as
a sigma model with $EAdS_3$ target space without taking the Liouville
limit. However, the derivation of the boundary action follows the
procedure in \cite{chvd}, and thus suffers from the same shortcomings
we discussed above. Moreover, our conclusions regarding the role of the
boundary action are rather different. In \cite{banadosritz}, on the basis
of the absence of local degrees of freedom in Chern-Simons theory, it is
claimed that asymptotically EAdS gravity is equivalent to the dynamics
of the boundary WZW model. As emphasised above, our present off-shell analysis
strongly suggests that this is not the case at the quantum level.
We agree with the analysis of Martinec \cite{martinec} of these issues
(but see \cite{kk0,kk2}).

The rest of the paper consists of Section 2 in which we present the
derivation and describe the significance of our boundary terms and Section
3 wherein the properties of these actions, in particular of the coset
WZW models, are described. In the Appendix we present our conventions
for three-dimensional gravity, Chern-Simons and WZW theories.

\section{Boundary Terms for Chern-Simons Gravity}

\subsection{Chern-Simons theory with gauge group $\gc$ or $G \times G$}

We will consider Chern-Simons theory with gauge group $G_{\CC}$, the
complexification of a (not necessarily compact) real Lie group $G$,
or with gauge group $G\times G$, on a three-manifold $M$ with boundary
$\partial M = \Sigma$.  We are interested in the odd (imaginary for $\gc$) 
part
\be
I_{CS}[A,\bar{A}]=I_{CS}[A]-I_{CS}[\bar{A}]
\ee
of the Chern-Simons action
\be
I_{CS}[A] = \int_M \tr(A\wedge dA + \frac{2}{3} A^3)\;\;,
\label{ics}
\ee
where, for $\gc$, $A$ and $\bar{A}$ are $\gc$-connections
related by hermitian conjugation (see Appendix A.1), and, for $G\times G$,
independent $G$-connections.

It is well known \cite{2+1witten,at}
that (modulo boundary terms, which will be the main focus of our
discussion later on) this action for $G=SU(2)$ or $G=SU(1,1)\sim SL(2,\RR)$ is 
(proportional to) the
Palatini action for Euclidean or Lorentzian three-dimensional
gravity with a positive or negative cosmological constant or, in other words,
(Euclidean) (anti-) de Sitter (E/A)dS gravity (see Appendix A.3).
This is summarized in the table below:
\be
\begin{array}{||l|c|c||}\hline\hline
 & \gc = SL(2,\CC) & G\times G \\ \hline \hline
G=SU(2) & \mbox{EAdS} & \mbox{EdS} \\ \hline
G=SU(1,1) & \mbox{dS} & \mbox{AdS} \\ \hline\hline
\end{array}
\ee
In each case the model target space geometry $(E/A)dS_3$ can be realised as
$\gc/G$ or $(G\times G)/G \sim G$ respectively, e.g.
$EAdS_3\sim SL(2,\CC)/SU(2)\sim H^3$ or $AdS_3\sim (SU(1,1)\times SU(1,1))/SU(1,1)
\sim SU(1,1)$, and $G$ is the (Euclidean or Lorentzian) tangent space group.

The even (real) part $I_{CS}[A]+I_{CS}[\bar{A}]$
of the Chern-Simons action 
gives rise to the so-called `exotic' action for three-dimensional
gravity \cite{2+1witten}, involving a Chern-Simons action for the spin connection.
Classically, for the vacuum Einstein equations, the ordinary
Palatini action and this action are equivalent. But when following
the usual minimal coupling prescription for matter, in the case of the
exotic action the torsion is determined by the energy-momentum and the
curvature by the spin-density, the opposite of what happens for the
Palatini action. For this reason, we will not consider this action in
the following.  

\subsection{$G$-invariance}

We now turn to a discussion of boundary terms appropriate for describing the
boundary dynamics of Chern-Simons gravity.  As discussed in the Introduction,  
it is natural to require that even on a manifold with boundary the action be
invariant under local Lorentz transformations (tangent space rotations) -
these are an artefact of the vielbein formulation of gravity and 
bulk Lorentz invariance should not interfere with the boundary degrees
of freedom. More generally this amounts to demanding strict invariance under
gauge transformations taking values in $G\subset \gc$ or in the 
diagonal subgroup $G\subset G\times G$.

$\gc$ or $G \times G$ gauge transformations act on $A$ and $\bar{A}$ as
(see Appendix A.1)
\be
A &\rightarrow& A^g \equiv g^{-1}A g + g^{-1}dg \\
\bar{A} &\rightarrow& \bar{A}^{\bar{g}}\equiv
\bar{g}^{-1}\bar{A} \bar{g} + \bar{g}^{-1}d\bar{g}\;\;,
\ee
with $g=\bar{g}$ iff $g$ takes values in $G$.  
The Chern-Simons action transforms as
\be
I_{CS}[A^g] = I_{CS}[A] 
 -\frac{1}{3}\int_M  \tr(g^{-1}dg)^3 + \oint \tr(A\we dg g^{-1})
\label{csag}
\ee
($\oint = \int_{\Sigma}$) 
and hence under $G$-transformations ($g=\bar{g}$)
the action $I_{CS}[A,\bar{A}]$ transforms
as
\be
I_{CS}[A^g,\bar{A}^g]=I_{CS}[A,\bar{A}]+\oint \tr (A-\bar{A})\wedge dg g^{-1}
\ee
(we will analyse the behaviour under general gauge transformations in Section 3.1). 
In particular, the WZ (winding number) term drops out.\footnote{Therefore there is no 
quantization condition on the Chern-Simons coupling constant - which is why we have
not worried about the normalization of the Chern-Simons action in (\ref{ics}).}
It is clear that this variation can be cancelled
by the gauge variation of the boundary term $\oint\tr (A\wedge \bar{A})$, because
\be
\oint\tr(A^g\wedge\bar{A}^g)= \oint\tr(A\wedge\bar{A})
+ \oint \tr (A-\bar{A})\wedge dg g^{-1} + \oint \tr(g^{-1}dg\wedge g^{-1}dg)
\ee
and the last term is a total derivative and hence zero.

Thus the most general $G$-invariant gravitational action takes the form
\be
I_{inv}[A,\bar{A}]=I_{CS}[A,\bar{A}] - \oint\tr(A\wedge \bar{A}) 
+ \oint {\cal F}(A,\bar{A})
\;\;,
\ee
where ${\cal F}(A,\bar{A})$ is a $G$-invariant functional of the boundary fields.
Excluding higher-derivative terms, such a functional can only be algebraic and
quadratic in $A-\bar{A}$.

We note in passing that the requirement of $G$-invariance can only be satisfied
in the special case $k=0$ of the general two-parameter family \cite{cscwitten}
of actions
\be
(k+is)I_{CS}[A] + (k-is)I_{CS}[\bar{A}] 
\ee
for $\gc$ (and likewise for $G\times G$), precisely the case we have been considering.
This is obvious from the above discussion because for a general linear combination
of Chern-Simons actions the Wess-Zumino (or winding number) term in (\ref{csag})
will not disappear
in the gauge variation, and this term cannot be cancelled by a {\em local} functional
of the boundary gauge fields.

This also has an explanation in the language of geometric
quantization. For $k\neq 0$ the prequantum line bundle is non-trivial,
and hence gauge transformations are necessarily implemented on wave
functions with a non-trivial cocycle. $G$-invariance of the action,
on the other hand, means that the $G$-action is implemented trivially
on the wave functions as $\psi(A)\ra\psi(A^g)$. Using the formalism of
\cite{cscwitten} it can indeed be checked that the prequantum operator
determined by our choice of boundary terms reduces to the standard
(trivial) generator of gauge transformations on functions of $A$ for
gauge transformations taking values in $G$.

\subsection{Boundary conditions and boundary terms}

It comes as a pleasant surprise that this choice of boundary terms, arising
from the requirement of $G$-invariance, is compatible
with the structure of asymptotically (EA)dS gravity where it is known that the
natural boundary conditions are such that either $A_z$ and $\bar{A}_{\zb}$ 
or $\azb$ and $\abz$ are fixed on the Euclidean boundary
(and likwise for AdS). Indeed, as we will now show,
these two requirements together uniquely determine the boundary terms.

The action appropriate for fixing either
$(\az,\abzb)$ or $(\azb,\abz)$ on the boundary is (cf.\ (\ref{icspm}))
\be
I^{\pm}[A,\bar{A}]&=&I^{\pm}_{CS}[A]-I^{\mp}_{CS}[\bar{A}]\nonumber\\
&=& I_{CS}[A,\bar{A}] \pm \oint \az\azb \pm \oint \abzb\abz\;\;.
\ee
Regardless of the boundary condition we denote the corresponding boundary value of 
the gauge field by $B=(B_z,B_{\bar{z}})$, so that for the two actions we have
\be
I^+: &&\az|_{\partial M}= B_z \quad \abzb|_{\partial M} = B_{\bar{z}}\\
I^-: && \abz|_{\partial M}= B_z \quad \azb|_{\partial M} = B_{\bar{z}} \;\;.
\ee
Note that, for $\gc$, the connection $B$, even though assembled from the $\gc$
connections $A$ and $\bar{A}$, is actually a $G$-connection, i.e.\ (see
Appendix A.1) it satisfies $B=\bar{B}$,
\be
B=\az dz + \abzb d\zb = \bar{B}
\ee
and likewise for $B=(\abz,\azb)$. 

The $G_{\CC}$ gauge transformations on $B$ inherited from the bulk gauge
transformations acting on $A$ and $\bar{A}$ is
\be
I^+: && B=\az dz + \abzb d\zb \ra \az^{g} dz + \abzb^{\bar{g}}d\zb\\
I^-: && B=\abz dz + \azb d\zb \ra \abz^{\bar{g}} dz + \azb^{g}d\zb\;\;,
\ee
e.g.
\be
B  = \az dz - (\az dz)^{\dagger} \ra \az^{g} dz - (\az^{g} dz)^{\dagger} \;\;.
\label{bg}
\ee
for $G=SU(2)$ (see Appendix A.1).  We will write this as
\be
B\ra B^{(g,\bar{g})}
\ee
and need to keep in mind that only 
for $g$ taking values in $G\subset G_{\CC}$
this reduces to the standard gauge
transformation $B\ra B^g$ of a $G$ gauge field,
\be
B^{(g,g)}=B^g\;\;.
\ee
This is precisely the extension of the action of the gauge group
$\G=\mbox{Map}(\Sigma,G)$ to $\GC$ on the space ${\cal A}$ of
$G$ gauge fields (with the quotient ${\cal A}/\GC$ essentially
the finite-dimensional moduli space of flat connections by the
Narasimhan-Seshadri theorem) that one also encounters in the geometric
quantization of $\gc$ Chern-Simons theory \cite{cscwitten} and that arises
here very concretely and naturally.

For the corresponding statements in the $G\times G$ case, see Section 3.3. 

To the action $I^{\pm}[A,\bar{A}]$
we are still free to add any (local) functional of the boundary 
gauge field $B$.  It is straightforward to see that there is precisely one such
term which makes the action $G$-invariant, i.e.\ compatible with the structure
of $I_{inv}[A,\bar{A}]$ we determined above, namely $\mp 2\oint\tr B_z B_\zb$.

Thus our desiderata
have uniquely determined the action which we will consider to be
\be
I^{\pm}_{tot}[A,\bar{A}]&=& I^{\pm}[A,\bar{A}]\mp 2\oint\tr B_z B_\zb\nonumber\\
                        &=& I_{CS}[A,\bar{A}] - \oint\tr(A\wedge \bar{A}) 
                        \pm \oint \tr(\az-\abz)(\azb-\abzb)\;\;.
\label{ibeta}
\ee
This shows that the two possibilities for ${\cal F}(A,\bar{A})$ compatible with the
boundary conditions are ${\cal F}(A,\bar{A})=\pm \oint \tr(\az-\abz)(\azb-\abzb)$
respectively. The above action, as we have written it, is imaginary for $\gc$.
We find it more convenient to keep this in mind than to clutter subsequent equations
with explicit factors of $i=\sqrt{-1}$.

As we will see, Chern-Simons theory with this choice of boundary
terms has several attractive features. In particular, by construction,
it is strictly $G$-invariant.  We can thus anticipate the appearance
of $\gc/G$ or $(G\times G)/G$ cocycles when studying the behaviour
of our gravitational action (or wave functions) under general gauge
transformations.

\subsection{Gravitational interpretation}

The above construction works for any $G$, but it is instructive to look at
the meaning of the boundary terms in the gravitational context or, for general
$G$, in the `gravitational parametrisation'
\be
A=\omega + \eta e \;\;,\;\;\;\; \bar{A}=\omega - \eta e\;\;,
\ee
where $\omega$ is a $G$-connection and $e$ is a one-form in the adjoint of $G$, 
which are, of course, to be identified 
with the spin-connection and dreibein in the gravitational context.
Here $\eta =i$ for $\gc$ (so that $A$ and $\bar{A}$ are related by hermitian
conjugation) and $\eta = 1$ for 
$G\times G$ (so that $A$ and $\bar{A}$ are independent $G$-connections). 
 
In terms of $(\omega,e)$, the Chern-Simons action is
\be
I_{CS}[A,\bar{A}] 
= 
4 \eta\int  \tr (e\wedge d\omega + e\wedge  \omega\wedge\omega  + \eta^2
\frac{1}{3}e^3) + 2\eta\oint_{\Sigma} \tr e\wedge \omega~.
\label{icsa}
\ee
Clearly the bulk part of the action (proportional to the Palatini action $I[e,\omega]$)
is invariant under gauge transformations taking values in $G\subset \gc$
or $G\subset G\times G$ since both $e$ and the curvature $F(\omega)=d\omega +
\omega\wedge\omega$
of $\omega$ transform homogeneously under $G$. The boundary term, however, is not.

One can recognize the terms that we have added to $I_{CS}[A,\bar{A}]$
as being proportional to the integral of the exterior curvature (essentially
one half of the Gibbons-Hawking term), 
\be
\oint \tr(A\we \bar{A})= 2\eta \oint \tr(e\we \omega) 
\ee
and an area term,
\be
\oint \tr(A_z - \bar{A}_z) (A_{\bar{z}} - \bar{A}_{\bar{z}})
=4 \eta^2 \oint \tr(e_z e_{\bar{z}}) 
\ee
Note first of all that the sign of the exterior curvature term is such
that it precisely cancels the exterior curvature boundary term 
in the action (\ref{icsa}).
Thus our action is just the sum of the Einstein-Palatini
action and the area term.  Since this action only depends on $e$ and
$F(\omega)$, both of which transform homogeneously under
$G$, this makes it manifest that the action is strictly
invariant under $G\subset\gc$, i.e.\ under $SU(2)$ or $SU(1,1)$ local
Lorentz transformations (frame rotations).

The sign of the area term is such that, for the appropriate choice
of boundary condition, for $G=SU(2)$ (EAdS) it {\em regularises}
the action in the sense that e.g.\ the classical action for $AdS_3$ is
zero. Thus the area term acts as a suitable counterterm \cite{bk,hs}
to provide finite conserved quantities from the Brown-York \cite{by}
prescription (modulo logarithmic divergences). Likewise for dS \cite{BdBM}.

Thus another way to arrive at the action we have been led to consider
is to demand that it be invariant
under Lorentz transformations (this rules out Gibbons-Hawking like terms)
and finite on classical asymptotically (EA)dS solutions. This fixes the
action uniquely up to Lorentz-invariant higher-derivative boundary terms 
(which do not contribute to the asymptotics in 3d \cite{bk}).

If we had added the two terms in (\ref{ibeta}) with the {\em opposite sign},
then the exterior curvature terms would have added up to the standard
Gibbons-Hawking term (or, rather, its Palatini counterpart $I_{GH}[e,\omega]$), 
giving an action with a well-defined
variational principle for the metric (or $e$) fixed on the boundary. 
The area term would still have regularized the classical action, 
since the Chern-Simons part of the action is finite all by itself, and
thus all that is required is a cancellation of divergences between 
the area and exterior curvature terms. Thus we would have obtained the 
geometrodynamics action with $e$ or the metric 
fixed on the boundary, with the appropriate counterterm $I_{ct}[e]$,
\be
&& I_{CS}[A] - I_{CS}[\bar{A}] + \oint \tr(A\we \bar{A}) 
\mp \oint \tr(\az - \abz)(\azb - \abzb)\nonumber\\ 
&\sim& I[e,\omega] + I_{GH}[e,\omega] +I_{ct}[e]\;\;.
\label{akk}
\ee
This is precisely the action Krasnov was led to investigate in the
Chern-Simons formulation of EAdS gravity in \cite{kk1}. However, for
reasons we will now explain we believe that the action he actually ends
up studying is the action (\ref{ibeta}) instead.\footnote{The author
of \cite{kk1} agrees with this conclusion \cite{kkpc}.}

One intriguing aspect of our choice of action, which will appear as a byproduct
of our general analysis in the next section, but which can also readily be seen
directly by using the Polyakov-Wiegman (PW) identities (\ref{pw}), 
is that on pure gauge configurations, $A=g^{-1}dg,
\bar{A}=\bar{g}^{-1}d\bar{g}$, it reduces to a WZW model (\ref{wzw})
with argument $g\bar{g}^{-1}$ (\ref{wzwgcg}).
This reflects the strict invariance of the action under
$G$-transformations, for which $g=\bar{g}$. 

The action (\ref{akk}) considered by Krasnov does not have this
invariance, as it contains the (non-invariant) Gibbons-Hawking term, and
indeed it can be checked that on-shell (on pure gauge transformations) the
classical action does not combine nicely into a single WZW model courtesy
of the PW identities. But precisely such a combination is claimed
to occur in \cite{kk1}, and this is a property of (\ref{ibeta}), not of 
the geometrodynamics action (\ref{akk}).

\section{The Boundary Action for Chern-Simons Gravity}

We find it notationally more convenient to treat the two cases $\gc$
and $G\times G$ separately, even though (as will be seen) they are exactly
analogous. We begin with $\gc$, deferring a discussion of $G\times G$
to Section 3.3.

\subsection{From Chern-Simons gravity to $G_{\CC}/G$ WZW models}

To establish the emergence of $\gc/G$ models from Chern-Simons gravity,
anticipated in various ways above, we now determine the behaviour of
$I^{\pm}_{tot}[A,\bar{A}]$ under a general $G_{\CC}$ gauge transformation.

First of all, note that gauge variation
of the Chern-Simons part $I^{\pm}[A,\bar{A}]$
of the total action is the sum of a chiral and an anti-chiral
WZW action (\ref{ipmg}),
\be
I^{+}[A^g,\bar{A}^{\bar{g}}]&=& I^{+}[A,\bar{A}] + (I^+[g,\az] + I^-[\bar{g},\abzb]) 
\nonumber\\
I^{-}[A^g,\bar{A}^{\bar{g}}]&=& I^{-}[A,\bar{A}] - (I^-[g,\azb] + I^+[\bar{g},\abz]) 
\;\;.
\ee
Usually at this stage \cite{chvd,ck} (after having gone `on-shell' in the bulk)
it is then argued that (on-shell on the boundary) the sum of these two
actions is equivalent to a non-chiral WZW action. In the present case, however,
this is much clearer and completely unambiguous since  
the boundary terms we have added to enforce $G$-invariance
precisely provide the relevant Polyakov-Wiegman
terms that allow us to combine these two chiral actions {\em off-shell}
into a single non-chiral action for the composite field $g\bar{g}^{-1}$.

Indeed a reasonably straightforward calculation, using the  PW identities (\ref{pw})
shows that 
\be
I_{tot}^{\pm}[A^g,\bar{A}^{\bar{g}}]=
I_{tot}^{\pm}[A,\bar{A}] \pm I^{\pm}[g\bar{g}^{-1},B]
\label{ipm}
\ee
where 
\be
I^{+}[h,B]
&=& I^{+}[h]
+ 2 \oint \tr 
(B_z \partial_{\bar{z}} h  h^{-1} - B_{\bar{z}}  h^{-1}\partial_z h ) +
2\oint \tr(B_{z}B_{\bar{z}} - B_{z} h B_{\bar{z}}h^{-1})\nonumber\\
I^{-}[h,B]
&=& I^{-}[h]
- 2 \oint \tr 
(B_z h^{-1}\partial_{\bar{z}} h   - B_{\bar{z}}  \partial_z h h^{-1}) +
2\oint \tr(B_{z}B_{\bar{z}} - B_{z}h^{-1} B_{\zb} h)\nonumber\\
&=& I^{+}[h^{-1},B] 
\ee
and where $I^\pm[h]$ is the WZW action (\ref{wzw}). We will discuss these actions
in more detail in the next section. 
 
Note that the above result is completely general: 

\begin{itemize}
\item
No assumptions have been made about the topology of $M$ or $\Sigma$, or about $\gc$. 
In particular, the results hold for $\Sigma$ the boundary of any region $M$
in space(-time), not only an `asymptotic' boundary.
\item
We did not need to make use of the bulk equations of motion to arrive at the boundary
action. In particular, no special coordinatization was required to turn some of the
components of the gauge field into Lagrange multipliers enforcing the bulk flatness
conditions.
\item 
We did not need to go on-shell for the boundary theory to combine two chiral WZW
models into a standard WZW model.
\end{itemize}

In all these aspects our derivation of the boundary action differs from
the standard derivations. We believe that our approach displays more clearly the
role of the boundary action, in particular in the full bulk quantum gravity 
theory. 

\subsection{Properties of the boundary action}

We list here some basic properties of the boundary actions (cocycles)
$I^{\pm}[h,B]$.
\begin{enumerate}
\item 
With dynamical boundary gauge fields $B$, 
the actions $I^{\pm}[h,B]$ with $h\in\gc$ are the actions of
diagonally gauged (thus anomaly free) $G_{\CC}/G$ WZW models.

The action we find is such that $h$ is restricted to 
take values in $\gc/G$, $h = g\bar{g}^{-1}$. Thus these are, despite appearance,
somewhat unusual WZW models.

In particular, since the cocycle $I^{\pm}[g\bar{g}^{-1},B]$
depends on $g$ and $\bar{g}$ only in the combination
$g\bar{g}^{-1}\in \gc/G$, it follows immediately that the gravitational
action $I_{tot}$ is invariant under gauge transformations with $g=\bar{g}$,
i.e.\ under local frame rotations, as expected, because
$I^{\pm}[h=1,B]=0$, 

Note, incidentally, that these ``$(\gc/G)/G$'' models can be shown to be
topological field theories, non-compact analogues of the more familiar
$G/G$ models \cite{gmodg}, which are of interest in their own right - 
these will be studied elsewhere.
\item
In our case, however, the $B$ are non-dynamical and these theories
are best thought of as WZW models coupled in a gauge-invariant way to 
external currents $B$. Setting these currents to zero, or evaluating
the gravitational action on classical solutions which are pure gauge,
$A=0^g=g^{-1}dg$, $\bar{A}=\bar{0}^{\bar{g}}$, one obtains the action
\be
I_{tot}^{\pm}[0^g,\bar{0}^{\bar{g}}]= \pm  
I^{\pm}[g\bar{g}^{-1},B=0] = \pm I^{\pm}[g\bar{g}^{-1}] \;\;.
\label{wzwgcg}
\ee
These WZW models are sigma models with target space $G_{\CC}/G$ and a
non-trivial (string theory) $B$-field $B_{NS}$. 
\item 
These actions (and their gauged counterparts) are (modulo an overall
factor of $i$) {\em real}, in contrast to ordinary (gauged) WZW models in 
Euclidean signature which have a complex action. (For example, the WZ term 
is real for unitary matrices but imaginary for (hermitian) matrices of the
form $g\bar{g}^{-1}$).
This is of course a consequence
of the fact that they arise as cocycles of the real gravitational action we have been
considering. Also, for $G$ compact they have a positive (semi-)definite kinetic
term and are thus well defined in Euclidean signature. Their continuation to Lorentzian
signature, however, would be {\em complex}, hence these theories are presumably
non-unitary in Minkowski space. 
\item
For $\gc = SL(2,\CC)$ the target space of the WZW model is the
three-dimensional space(-time) $(EA)dS_3$, precisely the three-manifold
whose quantum gravity we set out to describe. 
Thus, for any three-manifold $M$, the boundary action is a sigma model  
with target space $(EA)dS_3$, corresponding to the `on-shell' base space
$M$ of the Chern-Simons gravity theory. We will see that the same thing 
happens for $G\times G$.

This is reminiscent of the worldsheet for worldsheet construction \cite{greenws}
in $N=2$
string theory \cite{n2}. In that case one finds that the beta-function equations
(that is the conditions for that theory to be conformally invariant - quantum
consistent) of the worldsheet conformal field theory give rise to a target
space action, the field content and symmetry of which are identical to the
original worldsheet theory. In our case we find that the symmetries of our
three-dimensional base space theory give rise to a target space theory of the
same structure.

\begin{enumerate}
\item
In particular, for $G=SU(2)$, with the parametrisation 
\be
g\bar{g}^{-1} &=& \left (\begin{array}{cc}
e^\phi u\bar{u}+ e^{-\phi} & e^{\phi}u\\    
e^{\phi}\bar{u}  & e^{\phi}
\end{array}\right )\\
                g &=& \left (\begin{array}{cc}
e^{-\phi/2} & e^{\phi/2}u\\    
0 & e^{\phi/2}
\end{array}\right )
\ee
of the coset $SL(2,\CC)/SU(2)$, the action of the WZW model is
\be
I^+[g\bar{g}^{-1}] = 
2\int (\partial\phi\bar{\partial}\phi + e^{2\phi} \partial u
\bar{\partial} \bar{u}) \;\;.
\ee
This action has been studied in detail in the past e.g.\ in 
\cite{gawedzki,teschner}. It is also part of the string theory sigma model  
on $EAdS_3$ with non-zero NS B-field
\be 
ds^2 = d\phi^2 + e^{2\phi} dud\bar{u}, \quad B_{NS} = 
e^{2\phi} du\wedge d\bar{u}
\ee
as studied in detail e.g.\ in \cite{GKS,BORT,JMHO}. 
In \cite{teschner}
one sees clearly that the correlation functions are closely
related in construction and form to those of Liouville theory.
And in the limit that the string worldsheets are localized at radial infinity,
the worldsheet action reduces to Liouville theory \cite{d1d5}. 

\item
Likewise, for $G=SU(1,1)$ (this amounts to replacing 
$(u,\bar{u})\ra(iu,i\bar{u})$ in the above)
we find a target space with the $dS_3$ metric and an NS B-field, 
\be
ds^2 = d\phi^2 - e^{2\phi} dud\bar{u}, 
\quad B_{NS} = - e^{2\phi} du\we d\bar{u}.
\ee
This action can be reduced to Liouville theory by imposing further constraints
appropriate for asymptotically de Sitter gravity \cite{ck,bc}.
However, not much appears to be known about the properties of this sigma model
itself. 
An initial step in the study of this model could be a minisuperspace
analysis as carried out for the $SL(2,\CC)/SU(2)$ model in \cite{teschnermini}.
\end{enumerate}

\item In the path integral, the cocycle gives rise to wavefunction(al)s  
\be
\Psi^{\pm}[B] = \int D[A] D[\bar{A}] \ex{isI^{\pm}_{tot}[A,\bar{A}]},
\ee
($s$ is real for Euclidean $AdS_3$ and imaginary for Lorentzian
$dS_3$ \cite{cscwitten}) of the boundary gauge field $B$ 
which transform with a cocycle under $G_{\CC}$ gauge transformations,
\be
\Psi^{\pm}[B^{(g,\bar{g})}] = \ex{\pm is I^{\pm}[g\bar{g}^{-1},B]}\Psi^{\pm}[B]
\label{cocb}
\ee

\item
To understand the role of the $G_{\CC}/G$ sigma model (cocycle)
in the quantization
of Chern-Simons gravity, one can gauge fix the above path integral
by following the usual Faddeev-Popov procedure. Because of the non-trivial cocycle,
the integral over the gauge group will not factor out of the path integral.
Rather, one finds (see e.g.\ \cite{Carlip2}) 
that the gauge fixed version of the wave function factorizes 
into the standard gauge fixed Chern-Simons gravity
path integral $Z_{CS}[B]$ and the partition 
function of the WZW action $I^{\pm}[g^{-1}\bar{g},B]$, with the $B$ treated
as external sources,
\be
\Psi[B]=Z_{CS}[B]\times Z_{\gc/G}[J=B] \;\;.
\ee
In particular, the latter is a generating functional for correlation
functions of the left and right $h$-currents in the WZW model
(\ref{wzwgcg}) with target space $G_{\CC}/G$. The addition of Wilson
lines, gravitational sources, ending on the boundary will 
couple the bulk and boundary theories in a non-trivial way
and will lead to operator insertions in the boundary partition function. 
\end{enumerate}

\subsection{An analogous construction for $G\times G$}

As we noted before,
if the gauge fields $A$ and $\bar{A}$ are taken to be {\em independent} $G$ gauge
fields rather than complex conjugate $\gc$ gauge fields, 
then the (real) action $I_{CS}[A]-I_{CS}[\bar{A}]$ describes Lorentzian Anti-de Sitter
(ADS) gravity for $G=SU(1,1)$ or $G=SL(2,\RR)$, and Euclidean de Sitter (EdS) gravity
for $G=SU(2)$. 

Most of what we have done above for $\gc$ Chern-Simons gravity goes through verbatim 
in these cases as well with the replacements
\be
\gc \ra G \times G \;\;\;\;\;\; G\ra \mbox{diag}(G\times G)\;\;.
\ee
As EdS and AdS have some slightly different features, we treat them separately,
beginning with AdS.

For AdS we might be
interested in a Lorentzian boundary, and hence the boundary gauge field would be
\be
B=A_{\pm}dx^{\pm} + \bar{A}_{\mp} dx^{\mp}\;\;.
\ee
This is an $SU(1,1)$ or $SL(2,\RR)$ gauge field, $B^{\dagger}=
 -\sigma_{3}B\sigma_{3}$, and transforms as such under the diagonal subgroup
of $SU(1,1)\times SU(1,1)$. The two components of $B$ transform independently
under $SU(1,1)\times SU(1,1)$, and this extension of the $\G$-action on
$G$ gauge fields to a $\G\times\G$ action is the analogue in the present case
of the extension of the $\G$-action to $\GC$ encountered before. 

In this case $g$ and $\bar{g}$ are independent elements of $SU(1,1)$,
and $g\bar{g}^{-1}$ is thus an element of $(G\times G)/G \sim G$ itself.
The cocycle we obtain in this case is the standard $G=SU(1,1)$ or
$G=SL(2,\RR)$ WZW model which is real in Lorentzian signature.  Note that
once again the boundary action is thus a sigma model with target space the
three-manifold $G\sim AdS_3$ whose quantum gravity we set out to describe.
By imposing further asymptotically $AdS_3$
constraints, this model can be reduced to Liouville theory \cite{chvd}.
The coupling to the boundary gauge fields $B$ is such that with the latter
treated as dynamical the action is actually precisely that of the {\em
topological} $G/G$ model.  

For EdS, on the other hand, the boundary gauge field
\be
B=\az dz + \abzb d\zb
\ee
(say) is {\em not} an $SU(2)$ gauge field, $B^\dagger \neq -B$, because
$A$ and $\bar{A}$ are independent connections (we would 
have encountered a similar situation for dS gravity with a Lorentzian boundary).
Nevertheless, under the diagonal subgroup of $G \times G$ it transforms
as such and this is all that we will need. 

Consequently we are now choosing complex boundary conditions on the two independent
fields $\az$ and $\abzb$.  The gravitational action we are led to
with these boundary conditions and the requirement of $G$-invariance
is the sum of the standard Palatini action and an {\em imaginary} area
term. While this may appear strange from the gravitational point of view,
this is what emerges from our construction. However, these boundary conditions
are most likely not natural for EdS.

The cocycle is now an $SU(2)$ WZW model (in agreement with the fact that
$EdS_3 \sim S^3 \sim SU(2)$), which is complex in Euclidean space (and
unitary in Lorentzian signature).  With the $B$'s included as dynamical
fields, it is a topological $SU(2)/SU(2)$ model (with slightly unusual
reality properties).

\begin{center}
{\bf Acknowledgements}
\end{center}

We thank George Thompson for numerous helpful discussions.

The  work of GA and MO is supported in part by the European Community's
Human Potential Programme under contract HPRN-CT-2000-00131 Quantum
Spacetime. The research of MB is partially supported by EC contract
HPRN-CT-2000-00148.

\appendix

\section{Conventions}

\subsection{Lie algebras, gauge fields and gauge transformations}

Given a basis $\tau_{a}$ of the Lie algebra $\lg$ of $G$,
$\{\tau_{a},i\tau_{a}\}$ are a basis of $\lgc$.  A $G_{\CC}$ connection
$A$ can then be written as
\be
A = \omega + i e = \tau_a(\omega^a + i e^a) \;\;.
\ee
We also define the conjugate connection $\bar{A}$ by \cite{cscwitten}
\be
\bar{A} =\omega - i e = \tau_a(\omega^a - i e^a)
\ee
Thus $\bar{A}=-A^\dagger$ if the $\tau_{a}$ are anti-hermitian,
$\tau_{a}^{\dagger}=-\tau_{a}$, but not in general. In particular, 
$\bar{A}=A$ iff $A$ is a $G$-connection.

Likewise, for $G\times G$, we parametrise the two independent $G$-connections
$A$ and $\bar{A}$ as
\be
A &=& \omega +  e = \tau_a(\omega^a +  e^a) \nonumber\\
\bar{A} &=& \omega -  e = \tau_a(\omega^a - e^a) \;\;,
\ee
which satisfy $A^{\dagger}=-A$ and $\bar{A}^{\dagger}=-\bar{A}$ if the $\tau_a$
are anti-hermitian.

For $G=SU(2)$ or $G=SU(1,1)$ we have $G_{\CC} = SL(2,\CC) \sim SO(3,1)$ 
and $G\times G \sim SO(4)$ or $\sim SO(2,2)$. 
For $G=SU(2)$ we choose $\tau_a =-i\sigma_a/2$, $a=1,2,3$ ($\sigma_a$ are standard 
hermitian Pauli matrices), and thus $A^\dagger = -\bar{A}$, 
while for $G=SU(1,1)$ we take $\tau_a = (-i\s_3/2, \s_1/2, \s_2/2)$, $a=0,1,2$ with 
$A^\dagger = -\sigma_3 \bar{A} \sigma_3$. The $\tau_a$ satisfy
\be
{}[\tau_a,\tau_b]=\epsilon_{abc}\tau^c
\ee
where indices are raised and lowered with $\d_{ab}$ or
$\eta_{ab}=\mbox{diag}(-++)$ and we use the convention that
$\epsilon_{123}=\epsilon_{012}=1$. Explicitly, the curvature 
$F_A = dA + A^2$ of $A$ is
\be
F_{A}&=& \tau_{a}(d\omega^a + \frac{1}{2}\epsilon^{a}_{\;bc}\omega^b\wedge
\omega^c - \frac{1}{2}\epsilon^a_{\;bc}e^b\wedge e^c)\nonumber\\
     &+& i\tau_a (de^a + \epsilon^{a}_{\;bc}\omega^{b}\wedge e^c)
\label{FA}
\ee
for $\gc$, with an analogous expression for $G\times G$ obtained by
sending $e\ra -i e$.

$\gc$ gauge transformations act on $A$ and $\bar{A}$ as
\be
A &\rightarrow& A^g \equiv g^{-1}A g + g^{-1}dg \\
\bar{A} &\rightarrow& \bar{A}^{\bar{g}}\equiv
\bar{g}^{-1}\bar{A} \bar{g} + \bar{g}^{-1}d\bar{g}.
\ee
with   $\bar{g}^{-1} = g^\dagger$ for $G=SU(2)$, and
$\bar{g}^{-1} = \s_3 g^\dagger\s_3$  for $G=SU(1,1)$ 
respectively. 

Note that in both cases $g \ra \bar{g}$ is an (outer)
automorphism of $G_{\CC}$, $\overline{gh}=\bar{g}\bar{h}$, which
fixes $G\subset G_{\CC}$, i.e.\ $g=\bar{g} \Leftrightarrow g\in G$.
Thus $g\ra g\bar{g}^{-1}$ is the
projection from $G_{\CC}$ to the coset $G_{\CC}/G$.
In particular, for $G=SU(2)$ this is the projection $g\ra h= gg^{\dagger}$
onto (positive) hermitian matrices, $h^{\dagger}=h$.

Likewise, for the $G\times G$ theory, $g$ and $\bar{g}$ are independent $G$ gauge
transformations, and the map $g\ra g\bar{g}^{-1}$ is the
projection from $G\times G$ to $G$ with kernel the diagonal subgroup 
$G_d=\{(g,g)\in G\times G\}$.

\subsection{Chern-Simons and WZW actions}

The Chern-Simons action is
\be
I_{CS}[A]=\int\tr (A\wedge dA + \frac{2}{3}A^{3})\;\;.
\ee

Under a gauge transformation 
\be
A &\rightarrow& A^g \equiv g^{-1}A g + g^{-1}dg \\
\bar{A} &\rightarrow& \bar{A}^{\bar{g}}\equiv
\bar{g}^{-1}\bar{A} \bar{g} + \bar{g}^{-1}d\bar{g}.
\ee
the Chern-Simons action transforms as
\be
I_{CS}[A^g] = I_{CS}[A] + C[g,A]
\ee
where the cocycle $C[g,A]$ is
\be
C[g,A]= -\frac{1}{3}\int_M  \tr(g^{-1}dg)^3 + \oint \tr(A\we dg g^{-1})
\ee
and analogously for $I_{CS}[\bar{A}]$. 

The variation of $I_{CS}[A]$ is
\be
\d I_{CS}[A] = 2 \int\tr \d A \wedge F_{A} - \oint\tr A\wedge \d A\;\;.
\ee
Choosing a complex structure $J$ on the boundary, 
in terms of complex coordinates the boundary term reads
\be
\oint\tr A\wedge  \d A = \oint\tr (A_z \d A_\zb - A_\zb \d A_z)
\ee
where we introduce the convention that any expression of the form
$\oint {\cal O}_{z\zb}$ is to be interpreted as
\be
\oint {\cal O}_{z\zb}\equiv \oint dz\wedge d\zb \; {\cal O}_{z\zb}\;\;.
\ee
This makes it easy to switch between differential form and complex notation,
\be
\oint A\wedge B = \oint (A_z B_\zb - A_\zb B_z)\;\;,
\ee
but has the drawback that terms that look real are imaginary, 
and vice-versa, because $dz\wedge d\zb$ is imaginary,
\be
\overline{dz\wedge d\zb} =- dz\wedge d\zb \;\;.
\ee

To obtain an action that is suitable for fixing either $A_z$ or $A_\zb$ on the
boundary, we thus introduce the actions 
\be
I_{CS}^{\pm}[A]=I_{CS}[A]\pm\oint\tr A_z A_\zb
\label{icspm}
\ee
which have the on-shell variations
\be
\d I_{CS}^{+}[A]&=&2\oint\tr A_\zb \d A_z\nonumber\\
\d I_{CS}^{-}[A]&=&-2\oint\tr A_z \d A_\zb \;\;.
\ee
We can thus fix the boundary conditions
\be
A_z|_{\del M} = B_z \quad\quad\mbox{or}\quad\quad
A_\zb|_{\del M} = B_\zb 
\ee
for $I^+_{CS}[A]$ or $I^-_{CS}[A]$ respectively.

One can proceed analogously if one wishes to consider a boundary with a Lorentzian
metric, as for asymptotically AdS space-times. In this case, one replaces the
$(z,\bar{z})$ by light-cone coordinates $x^{\pm}$, with $dx^+\wedge dx^-$ real.

Under a gauge transformation, one finds that $I^{\pm}_{CS}[A]$ transforms as
\be
I_{CS}^{+}[A^g]&=& I_{CS}^{+}[A] + I^+[g,B_z] \\
I_{CS}^{-}[A^g]&=& I_{CS}^{-}[A] - I^-[g,B_\zb] \;\;,
\label{icspmg}
\ee
where
\be
I^+[g,B_z] &=& I^+[g] + 2 \oint\tr B_z \del_{\zb}g g^{-1}\nonumber \\
I^-[g,B_\zb] &=& I^-[g] + 2\oint \tr B_\zb \del_z g g^{-1}\label{ipmg}
\ee
with $I^{\pm}[g]$ the WZW actions 
\be
I^{\pm}[g]\equiv  \oint \tr(g^{-1}\partial_z g g^{-1} \partial_{\bar{z}} g)
\mp \frac{1}{3}\int_M \tr(g^{-1}dg)^3\;\;.
\label{wzw}
\ee
They satisfy the Polyakov-Wiegman (PW) identities
\be
I^{+}[gh] &=& I^{+}[g] + I^{+}[h] 
          + 2 \oint\tr  g^{-1}\partial_{z} g \partial_{\bar{z}}h h^{-1}\nonumber\\
I^{-}[gh] &=& I^{-}[g] + I^{-}[h] 
          + 2 \oint\tr  g^{-1}\partial_{\bar{z}} g \partial_{z}h h^{-1}\;\;.
\label{pw}
\ee
To make contact with the usual way of writing the WZW action, note that
with $d^2z=|dz\wedge d\zb|$ we have
\be
-i I^{\pm}[g]= -\oint d^2z\tr(g^{-1}\partial_z g g^{-1} \partial_{\bar{z}} g)
\pm \frac{i}{3}\int_M \tr(g^{-1}dg)^3
\ee

\subsection{$SL(2,\CC)$ Chern-Simons gravity}

We review the Chern-Simons formulation of EAdS and dS
gravity. We omit the discussion for AdS and EdS gravity which is
precisely analogous.

In the first order formalism the fields
are a dreibein (orthonormal frame) $e^a$ and the spin connection
$\omega^a_{\;b}$. We use the convention that $a=1,2,3$ for $EAdS_3$
and $a=0,1,2$ for $dS_3$.

The torsion-free condition for the spin-connection is 
\be
de^a + \omega^{a}_{\;b}\wedge e^b = 0\;\;.
\ee
The curvature two-form is 
\be
R^{a}_{\;b} = d\omega^{a}_{\;b}+\omega^{a}_{\;c}\wedge \omega^{c}_{\;b}\;\;.
\ee
In three dimensions, the Einstein equations with 
negative (positive) cosmological constant are equivalent to the constant curvature
condition
\be
R^{a}_{\;b}=\mp e^{a}\wedge e_{b}\;\;,
\ee
where indices are raised and lowered with the orthonormal (Euclidean or Minkowskian)
tangent space metric $\delta_{ab}$ or $\eta_{ab}=\mbox{diag}(-++)$.
Introducing the curvature radius by scaling $e^a\ra e^a/\ell$, 
the cosmological constant could be shifted away from its canonical value
$\Lambda = \pm 1$.

To establish the connection of these equations with the equations of motion of
Chern-Simons theory, one introduces the dual spin connection
\be
\omega^a = \mp \frac{1}{2}\epsilon^{abc}\omega_{bc}
\ee
where $\epsilon_{123}=1$, $\epsilon_{012}=1$ 
and the upper (lower) sign refers to $EAdS_3$ ($dS_3$) respectively.
Then the equations become
\be
&&de^a + \epsilon^{a}_{\;bc}\omega^{b}\wedge e^c=0\nonumber\\ 
&& d\omega^a + \frac{1}{2}\epsilon^{a}_{\;bc}\omega^b\wedge \omega^c =  
\frac{1}{2}\epsilon^{a}_{\;bc}e^b \wedge e^c\;\;.
\ee
Comparing with (\ref{FA}), we see that these are precisely the equations
$F_{A}=0$ of the $G_{\CC}=SL(2,\CC)$-connection $A$ for $G=SU(2)$ or $G=SU(1,1)$.
 
The relation between the Chern-Simons action and
the Palatini action for three-dimensional (EA)dS gravity, 
\be
I[e,\omega] = \int e^a\wedge (d\omega + \frac{1}{2}\epsilon_{abc}\omega^b\wedge\omega^c) -
\frac{1}{6}\epsilon_{abc}e^a\wedge e^b\wedge e^c~,
\ee
is
\be
I_{CS}[A] - I_{CS}[\bar{A}] = \mp 2i(I[e,\omega] \mp \half \oint e^a\we \omega )
\label{csaab}
\ee
(the signs arise beause for $G=SU(2)$ we are raising and lowering indices with
$\d_{ab}$, not with $\tr\tau_{a}\tau_{b} \sim -\d_{ab}$).
This action thus descibes Euclidean gravity with a negative cosmological constant
for $G=SU(2)$, and Lorentzian gravity with a positive cosmological constant
for $G=SU(1,1)$. 

\newpage

\rnc{\Large}{\normalsize}


\begin{thebibliography}{00}
\addcontentsline{toc}{section}{References}
\frenchspacing
\small
\addtolength{\itemsep}{-4pt}

\bibitem{bh} J.D. Brown and M. Henneaux, ``Central charges in the
canonical realization of asymptotic symmetries: an example from
three-dimensional gravity'', Comm. Math. Phys.104(1986) 207.

\bibitem{park1} M. Park, ``Statistical Entropy of Three-dimensional Kerr-De Sitter
Space'', Phys. Lett. B440 (1998) 275-282, {\tt hep-th/9806119}.

\bibitem{dscft1} 
A. Strominger,``The dS/CFT correspondence'', JHEP 0110:034, 2001, {\tt
hep-th/0106113}; 
M. Spradlin, A. Strominger, A. Volovich, ``Les Houches Lectures on De
Sitter Space'', {\tt hep-th/0110007};

\bibitem{GKS}
A. Giveon, D. Kutasov and N. Seiberg,  ``Comments on string theory on
$AdS_3$'', Adv. Theor. Math. Phys. 2 (1998) 733, {\tt
hep-th/9806194}; D. Kutasov, N. Seiberg, ``More Comments on String Theory on $AdS_3$'',
JHEP 9904 (1999) 008, {\tt hep-th/9903219}; A. Giveon, D. Kutasov, ``Notes on $AdS_3$'',
Nucl. Phys. B621 (2002) 303-336, {\tt hep-th/0106004}. 

\bibitem{BORT} J. de Boer, H. Ooguri, H. Robins, J. Tannenhauser,
``String Theory on $AdS_3$'', JHEP 9812 (1998) 026, {\tt hep-th/9812046}.

\bibitem{JMHO}
J. Maldacena and H. Ooguri, ``Strings in $AdS_3$ and $SL(2,\RR)$
WZW model.1: The Spectrum'', J. Math. Phys. 42 (2001) 2929-2960,
{\tt hep-th/0001053};  J. Maldacena, H. Ooguri and J. Son,
``Strings in $AdS_3$ and the $SL(2,\RR)$ WZW model.  Part 2: The
euclidean black hole'', J. Math. Phys. 42 (2001) 2961-2977,
{\tt hep-th/0005183}; J. Maldacena and H. Ooguri, ``Strings in $AdS_3$
and the $SL(2,\RR)$ WZW model.  Part 3: Correlation functions'', 
Phys. Rev. D65 (2002) 106006, {\tt hep-th/0111180}.

\bibitem{2+1witten}
E. Witten, ``2+1 dimensional gravity as an exactly soluble system'', Nucl.
Phys. B311 (1988) 46.

\bibitem{at} 
A. Achucarro and  P.K. Townsend, ``A Chern-Simons action for
three-dimensional anti-de sitter supergravity theories'', Phys. Lett. B180 (1986) 89.

\bibitem{carlip} For a review see: S. Carlip, ``Quantum Gravity in 2+1 Dimensions'',
Cambridge University Press (1998).

\bibitem{wittencs}
E. Witten, ``Quantum field theory and the Jones polynomial'', Commun. Math. Phys.
121 (1989) 351.

\bibitem{banados1} M. Banados, ``Global Charges in Chern-Simons theory and the 2+1
black hole'', Phys. Rev. D52 (1996) 5816, {\tt hep-th/9405171}.

\bibitem{park2} P. Oh and M. Park, 
``Symplectic Reduction and Symmetry Algebra in Boundary Chern-Simons theory'',
Mod. Phys. Lett. A14 (1999) 231-238, {\tt  hep-th/9805178}; M. Park, ``Symmetry
Algebras in Chern-Simons Theories with Boundary: Canonical Approach'', Nucl. Phys. B544
(1999) 377-402, {\tt hep-th/9811033}.

\bibitem{EMSS}
G. Moore, and N. Seiberg, ``Taming the conformal zoo'', Phys.Lett. B220 (1989) 422;
S. Elitzur, G. Moore, A. Schwimmer and N. Seiberg, ``Remarks on the
canonical quantization of the Chern-Simons-Witten theory'', Nucl. Phys.
B326 (1989) 108-134.

\bibitem{chvd}
O. Coussaert, M. Henneaux and P. van Driel, ``The asymptotic dynamics of
three-dimensional Einstein gravity with a negative cosmological
constant'', Class.Quant.Grav. 12 (1995) 2961-2966, 
{\tt gr-qc/9506019}; M. Henneaux, L. Maoz and A. Schwimmer,
``Asymptotic dynamics and asymptotic
symmetries of three dimensional extended AdS supergravity'', Annals
Phys. 282 (2000) 31-66, {\tt hep-th/9910013}. 

\bibitem{ksss}
K. Skenderis, S.N. Solodukhin, ``Quantum effective action from the AdS/CFT
correspondence'', Phys.Lett. B472 (2000) 316-322, {\tt  hep-th/9910023}.

\bibitem{bers}
 K. Bautier, F. Englert, M. Rooman, Ph. Spindel, ``The Fefferman-Graham Ambiguity and
AdS Black Holes'', Phys.Lett. B479 (2000) 291-298, {\tt hep-th/0002156}.

\bibitem{kk0} 
K. Krasnov, ``3D Gravity, Point Particles and Liouville Theory'',
Class.Quant.Grav. 18 (2001) 1291-1304, {\tt hep-th/0008253}.

\bibitem{kv} D. Klemm, L. Vanzo, ``De Sitter Gravity and 
Liouville Theory'', {\tt hep-th/0203268}.

\bibitem{d1d5}
N. Seiberg and E. Witten, ``The D1/D5 system and singular CFT'',
JHEP 9904 (1999) 017, {\tt hep-th/9903224}.

\bibitem{martinec} 
E. Martinec, ``Matrix Models of AdS Gravity'', {\tt hep-th/9804111}; E. Martinec,
``Conformal Field Theory, Geometry, and Entropy'', {\tt hep-th/9809021}

\bibitem{kk2}
K. Krasnov, ``$\Lambda<0$ Quantum Gravity in 2+1 Dimensions I: Quantum States and Stringy
S-Matrix'', Class. Quant. Grav. 19 (2002) 3977-3998, {\tt hep-th/0112164}.

\bibitem{ck}
S. Cacciatori and D. Klemm, ``The asymptotic dynamics of de sitter gravity
in three-dimensions'', Class. Quant. Grav. 19 (2002) 579-588, {\tt
hep-th/0110031}.

\bibitem{cscwitten}
E. Witten, ``Quantization of Chern-Simons Gauge Theory with complex
gauge group'', Commun. Math. Phys. 137 (1991) 29-66.

\bibitem{gmodg}
M. Spiegelglas and S. Yankielowicz, 
``$G/G$--Topological Field Theories by Cosetting $G_k$'',
Nucl. Phys. B393 (1993) 301, {\tt hep-th/9201036}; 
M. Blau and G. Thompson, ``Derivation of the Verlinde Formula from Chern-Simons Theory
and the $G/G$ model'', Nucl. Phys. B408 (1993) 345-390, {\tt hep-th/9305010}.

\bibitem{greenws}
M. B. Green, ``World sheets for world sheets'', Nucl. Phys. B293 (1987) 593-611;
``World-volumes and string target spaces'', Fortsch. Phys. 44 (1996) 551-563;
Nucl.Phys.Proc.Suppl. 49 (1996) 123-132, {\tt hep-th/9602061}.

\bibitem{n2} D. Kutasov and E. Martinec, ``New Principles for
String/Membrane Unification'', Nucl.Phys. B477 (1996) 652-674, {\tt hep-th/9602049}.

\bibitem{kk1}
K. Krasnov, ``On holomorphic factorization in
asymptotically AdS 3-D gravity'', {\tt hep-th/0109198}.

\bibitem{banadosritz}
M. Banados and A. Ritz, ``A note on classical string dynamics on $AdS_3$'',
Phys. Rev. D60 (1999) 126004, {\tt hep-th/9906191}.

\bibitem{banadosothers}
M. Banados, T. Brotz, M. Ortiz, ``Boundary dynamics and the statistical mechanics of 
the 2+1 dimensional black hole'', Nucl. Phys. B545 (1999) 340-370, {\tt
hep-th/9802076}; M. Banados, ``Three-dimensional quantum geometry and black holes'',
Invited talk at 2nd La Plata Meeting on Trends in
Theoretical Physics, Buenos Aires, Brazil, 28 Nov - 4 Dec 1998, {\tt hep-th/9901148};
M. Banados, O. Chandia, A. Ritz, ``Holography and the Polyakov action'', Phys. Rev. D65
(2002) 126008, {\tt hep-th/0203021}. 

\bibitem{bk} 
V. Balasubramanian and P. Kraus, ``A stress tensor for anti-de Sitter
gravity'', Commun. Math. Phys. 208 (1999) 413, {\tt hep-th/9902121}.

\bibitem{hs} 
M. Henningson and K. Skenderis, ``The holographic Weyl anomaly'', 
JHEP 9807(1998) 023, {\tt hep-th/9806087}.

\bibitem{by}
J.D. Brown and J.W. York, ``Quasilocal energy and conserved charges
derived from the gravitational action'', Phys. Rev. D47 (1993) 1407.

\bibitem{BdBM}
V. Balasubramanian, J. de Boer and D. Minic, ``Mass, entropy and holography in 
asymptotically de Sitter spaces'', Phys. Rev. D65 (2002) 123508, {\tt hep-th/0110108}.

\bibitem{kkpc} K. Krasnov, private communication

\bibitem{gawedzki}
K. Gawedzki and A. Kupiainen, ``Coset construction from functional
integrals'', Nucl. Phys. 320 (1989) 625; K. Gawedzki, ``Non-compact
WZW conformal field theories'', in: Proceedings of NATO ASI Cargese
1991, eds. J. Fr\"ohlich, G. 't Hooft, A. Jaffe, G. Mack, P.K. Mitter,
R. Stora (Plenum Press, 1992) 247-274, {\tt hep-th/9110076};
K. Gawedzki, ``CFT, a case study'', {\tt hep-th/9904145}.

\bibitem{teschner}
J. Teschner, ``On the structure constants and fusion rules in the
$SL(2,\CC)/SU(2)$ WZW model'', Nucl. Phys. B546 (1999) 390-422, {\tt
hep-th/9712256}.

\bibitem{bc} B.C. da Cunha, ``Dwelling on de Sitter'', {\tt hep-th/0208018}. 

\bibitem{teschnermini}
J. Teschner, ``The minisuperspace limit of the $SL(2,\CC)/SU(2)$ WZNW 
model'', Nucl. Phys. B546 (1999) 369-389, {\tt hep-th/9712258}.

\bibitem{Carlip2} S. Carlip, ``Inducing Liouville theory from 
topologically massive gravity'', Nucl. Phys. B362 (1991) 111-124.

\end{thebibliography}
\end{document}